\documentstyle[preprint,aps]{revtex}

\begin{document}
\draft
\preprint{FNUSAL - 2/95}

\title{A Consistent Study of the
the Low Energy Baryon Spectrum
and the Nucleon-Nucleon Interaction
within the Chiral Quark Model}

\author{A. Valcarce $^{(1)}$, P. Gonz\'alez $^{(2)}$,
F. Fern\'andez$^{(1)}$ and V. Vento $^{(2)}$}

\address{$(1)$ Grupo de F\' \i sica Nuclear \\
Universidad de Salamanca, E-37008 Salamanca, Spain}

\address{$(2)$ Departamento de F\' \i sica Te\'orica e
Instituto de F\' \i sica Corpuscular \\
Universidad de Valencia-C.S.I.C, E-46100
Burjassot, Valencia, Spain}

\maketitle

\begin{abstract}
By solving the Schr\"{o}dinger equation for the three-quark system in
the hyperspherical harmonic approach,
we have studied the low energy part of the nucleon and $\Delta$ spectra
using a quark-quark interaction which reproduces the
nucleon-nucleon phenomenology. The quark-quark hamiltonian considered
includes, besides the usual one-gluon exchange,
pion and sigma exchanges generated by the
chiral symmetry breaking. The baryonic spectrum
obtained is reasonable and the resulting
wave function gives consistency to
the ansatz used in the two baryon system.
\end{abstract}

\narrowtext
\newpage

An exciting challenge in contemporary nuclear
physics is to achieve a
unified description of the baryon spectrum and
the baryon-baryon interaction.
Models have been widely used to study the properties of the
hadron spectrum due to the impossibility to solve
Quantum Chromodynamics (QCD) at the
current moment.
In particular, the quark potential models incorporate
the perturbative one-gluon exchange quark-quark ($qq$) potential
($V_{\rm OGE}$) derived from QCD as well as a parametrization
of some nonperturbative effects through a $qq$ confining
potential ($V_{\rm con}$) \cite{RUJ}. Such an {\em effective
theory} [the quark-gluon coupling constant is taken as an
effective one and the constituent quark masses ($m_q$) are parameters
fitted from the baryon magnetic moments] provides a reasonable
understanding of the baryonic spectrum and the baryon
properties \cite{ISG}.
The adequate description of the low energy baryonic
spectrum comes in part from the
fact that the OGE interaction has an
appropriate spin-isospin dependence
which is responsible for the hyperfine
splitting (e.g. $\Delta$-N mass
difference).

The idea of an interquark potential has been also
used to study baryon-baryon interactions.
Actually, the repulsive core of the nucleon-nucleon (NN)
force has been shown to arise from the color-spin structure
of the $V_{\rm OGE}$ \cite{FAE}. Nevertheless, the same
scheme has been proven incapable of describing the long
range part of the NN interaction unless pion exchange
between quarks is introduced. From the basic theory,
the origin of the one-pion exchange potential ($V_{\rm OPE}$)
is associated with the spontaneous chiral symmetry breaking
of QCD. Moreover, the inclusion of the
$qq$ sigma potential ($V_{\rm OSE}$) consistently with its
chiral partner (the pion) allows also to reproduce the
intermediate range NN interaction as well as the
deuteron properties \cite{FER,VAL}.

The consistent description of both scenarios (baryon
spectrum and nucleon-nucleon interaction) in a model based on
interactions between quarks is still an open problem.
There are different spectroscopic models in the
literature \cite{ISG} that through a fit of the parameters (sometimes
an independent fit for the positive and negative parity sectors)
are able to describe the low-energy baryon spectrum.
Other quark models have been
successfully used to reproduce the
nucleon-nucleon interaction \cite{FAE}.
Certainly, to judge critically
a model one should study the predictions in both
sectors.

We start from a $qq$ interaction
whose parameters are determined
studying the two nucleon
properties and charge exchange reactions, and then, we proceed to
analyze the baryon spectrum for this interaction. As we
shall discuss, the predictions for the baryon masses and
the baryon wave functions are not only a stringent
test of the interaction but also a consistency test of the
ansatz wave function used to generate the NN interaction
in the resonating group method (RGM) calculations.

The quark-quark potential is explicitly given by\cite{FER},

\begin{equation}
V_{qq} (\vec{r}_{ij}) = V_{con} (\vec{r}_{ij}) +
V_{OGE} (\vec{r}_{ij})
+ V_{OPE} (\vec{r}_{ij})  + V_{OSE} (\vec{r}_{ij}) \, ,
\end{equation}

\noindent
where $\vec{r}_{ij}$ is the interquark distance $(\vec{r}_{ij} =
\vec{r}_{i} - \vec{r}_{j})$.
$V_{OGE}$ is the one-gluon exchange potential
with a smeared (of $r_0$ range) $\delta$ function term in order to avoid
an unbound spectrum \cite{BHA}:

\begin{equation}
V_{OGE} ({\vec r}_{ij}) =
{1 \over 4} \, \alpha_s \, {\vec
\lambda}_i \cdot {\vec \lambda}_j
\Biggl \lbrace {1 \over r_{ij}} -
{1 \over {4 \, m^2_q}} \, \biggl [ 1 + {2 \over 3}
{\vec \sigma}_i \cdot {\vec
\sigma}_j \biggr ] \,\,
{{e^{-r_{ij}/r_0}} \over
{r_0^2 \,\,r_{ij}}}
- {1 \over {4 m^2_q \, r^3_{ij}}}
\, S_{ij} \Biggr \rbrace \, ,
\end{equation}

\noindent
$\alpha_s$ is the effective quark-quark-gluon coupling constant and the
$\lambda 's$ are the $SU (3)$ color matrices. The $\sigma ' s$
stand for the spin Pauli matrices and $S_{ij}$ is the quark tensor operator
$S_{ij} = 3 (\vec{\sigma}_i \, . \, \hat{r}_{ij}) (\vec{\sigma}_j
\, . \, \hat{r}_{ij}) - \vec{\sigma}_i \, . \vec{\sigma}_j $.

\noindent
$V_{OPE}$ and $V_{OSE}$ are the one-pion
and one-sigma exchange potentials given by:

\begin{eqnarray}
V_{\rm OPE} ({\vec r}_{ij}) & = & {1 \over 3}
\, \alpha_{ch} {\Lambda^2  \over \Lambda^2 -
m_\pi^2} \, m_\pi \, \Biggr\{ \left[ \,
Y (m_\pi \, r_{ij}) - { \Lambda^3
\over m_{\pi}^3} \, Y (\Lambda \,
r_{ij}) \right] {\vec \sigma}_i \cdot
{\vec \sigma}_j + \nonumber \\
 & & \left[ H( m_\pi \, r_{ij}) - {
\Lambda^3 \over m_\pi^3} \, H( \Lambda \,
r_{ij}) \right] S_{ij} \Biggr\} \,
{\vec \tau}_i \cdot {\vec \tau}_j \, ,
\end{eqnarray}

\begin{equation}
V_{\rm OSE} ({\vec r}_{ij}) = - \alpha_{ch} \,
{4 \, m_q^2 \over m_{\pi}^2}
{\Lambda^2 \over \Lambda^2 - m_{\sigma}^2}
\, m_{\sigma} \, \left[
Y (m_{\sigma} \, r_{ij})-
{\Lambda \over {m_{\sigma}}} \,
Y (\Lambda \, r_{ij}) \right] \, ,
\end{equation}

\noindent
where $m_\pi$ ($m_\sigma$) is the pion (sigma) mass,
$\alpha_{ch}$ is the chiral coupling constant related to the
$\pi NN$ coupling constant by
$\alpha_{ch} =
\left( 3 \over 5 \right)^2
{ g_{\pi NN}^2 \over {4 \pi}} { m_{\pi}^2
\over {4 m_N^2}}$,
$\Lambda$ is a cutoff
parameter, and $Y(x), H(x)$ are the Yukawa functions defined as:

\begin{equation}
Y (x) =  \frac{e^{-x}}{x} \, \, \, \, \, , \, \, \, \, \,
H (x) =  (1 + \frac{3}{x} + \frac{3}{x^2}) Y (x)
\end{equation}

With this interaction, the two baryon system has been studied in
the framework of the RGM.
For the spatial part of the
wave function of the quarks, a harmonic
oscillator ground state was
assumed,

\begin{equation}
\eta_{\rm os} (\vec{r}_i - \vec{R} \, ) =
\left( {1 \over {\pi b^2}} \right)^{3/4} e^{-(\vec{r}_i
- \vec{R}\,)^2/ 2b^2} \, ,
\end{equation}

\noindent
where $\vec{R}$ is a parameter which
determines the position of the baryon and
$b$ is the harmonic oscillator constant.

Using a unique set of parameters, the NN scattering phase
shifts \cite{FER}, the static and electromagnetic
properties of the deuteron \cite{VAL}
and the charge-exchange reaction  $p p
\rightarrow n \Delta^{++}$ \cite{FNU} have been
well reproduced. In fact, the OPE and OSE
provide the medium and
long-range behavior required to reproduce
the NN phenomenology.
Additionally, the pure nucleon-$\Delta$
interaction has the correct
properties as indirectly derived from
the $\pi NN$ system \cite{ANU}.
Let us note, that even reproducing
all these observables there is
no constraint on the radial structure of the confining
potential, since the baryon-baryon
interaction does not depend on it \cite{SHI}.

To evaluate the baryonic spectrum with the just described
potential ($V_{con} + V_{OGE} + V_{OPE} + V_{OSE}$)
we have solved
exactly the Schr\"{o}dinger equation in
the hyperspherical harmonic
approach \cite{BAL}, this is by expanding
the three-quark wave function in terms of
a set of basis functions on the hypersphere
unity (of dimension 6, the number of
independent coordinates). For the potentials
we make use of, the truncation
of the series (for the positive parity
states we have included the first two terms
in the hyperspherical expansion, k=0 and
k=2, and for the negative parity states only
the k=1 term) does not suppose any significant
loss of precision \cite{DES}. As we
have no coupling to continuum channels our states
do not exhibit resonance features.

As stated above, the radial structure of the confinement
cannot be fixed from the NN interaction. However, the
baryon spectrum strongly depends on the confining potential.
In fact, the widely used quadratic potential places the
Roper resonance 1 GeV above the ground state (if the
strength is fixed variationally as in RGM), while it
appears in the correct position if a linear potential is
used. This fact favors
a linear form for the confinement
[$V_{con} ({\vec r}_{ij}) = - a_c \,
(\vec{\lambda}_i \cdot \vec{\lambda}_j ) \, r_{ij}$]
as suggested by meson spectroscopy and
lattice calculations \cite{GUT},
to which we shall restrict from now on.
The values of the parameters are taken from
Ref. \cite{VAL} and quoted in Table I,
and the predicted $N$ and $\Delta$ spectra appear in
Figure 1.

It is worth to note that the chiral pseudoscalar
interaction contributes (together with the OGE)
to the hyperfine splitting,
driving the value of $\alpha_s$ to 0.4 - 0.5, close to the QCD one.
On the other hand, the $(\vec{\sigma} . \vec{\sigma}) (\vec{\tau} .
\vec{\tau})$ structure
of the pion interaction gives
attraction for symmetric spin-isospin pairs and
repulsion for antisymmetric ones (a quite distinctive
feature since the color-magnetic part of the OGE
gives similar contributions in both cases).
This lowers
the position of the first nucleon radial excitation
[Roper resonance, $N (1440)$] with regard to the
first negative parity state
solving part of the discrepancy between usual
two-body potential models (the
predicted relative energy positions
of the Roper and the first negative parity
appear inverted) and experiment. To have the
same effect (quasidegeneration of the first
positive and negative parity excitations)
in a model without pion interaction, the
OGE exchange Coulomb interaction has to be
very much strengthed \cite{DES}.
A complete inversion as required by
the experimental data can be achieved just by a small
modification of the values of the pion interaction
constants. For instance, using $\Lambda = 5.0 \,
fm^{-1}$ instead of $\Lambda = 4.2 \, fm ^{-1}$
one obtains 516 MeV for the excitation energy of
the Roper resonance and 534 MeV for the first negative
parity excited state. Nevertheless this modification
would destroy the agreement in the NN system.

Quantitatively, the relevant role played by
the pion interaction to this
respect is shown in Table II where the results
with the full potential [$N (1440)$ and $N^* (1535)$
almost degenerated]
are compared to the ones obtained when the pion
interaction is removed and
the OGE strength is fitted to
the experimental $\Delta -N$ mass difference. As can
be seen in this table, with the suppression of the pionic
interaction, the $N^* \, (1535)$
appears deeply bound with respect to the Roper resonance.

Regarding the negative parity states of the nucleon and $\Delta$,
they appear almost degenerated
due to the potential we use.
The only difference
comes from the small contribution of
the pion tensor interaction, which
depends on the total angular momentum $J$.
On the other hand, the splitting among the
negative parity states is insufficient
as compared to data. This might be an indication
of the need of fine tuning the
spin-spin interaction (and correspondingly a
refitting of the parameters in
the NN sector) as well as a signature of
the possible relevance of spin-orbit
terms, not considered here. Let us
point out that due to the
value of our coupling constant $\alpha_s$
and to the fact that the Galilei invariant
spin-orbit force of the one-sigma exchange
cancels the corresponding term  of
the one gluon exchange \cite{SPO} the spin-orbit
strength is very much reduced as compared to
other models \cite{ISG} where its inclusion
would destroy the overall fit of the
spectrum.
Although a refinement of the potential
incorporating a spin-orbit contribution
has been recently proposed \cite{SPO}
the existence of Galilei non-invariant terms,
spin-orbit force associated to confinement,
relativistic corrections at the
same order,... prevent making any definitive
statement about how relevant the
spin-orbit interaction should be. As
we are interested in
in the consistent treatment of the
one and two body problems,
we do not proceed further
along this direction and
center our attention in the examination of the wave functions.

In Figure 2 we compare the wave function
we obtain
for the nucleon ground state with the
ansatz used in the RGM formalism
for three different values of the oscillator
parameter $b$. Although at long
range all the wave functions look very similar, the
discrepancy at the origin
reaches a factor four. This effect may be very
important for quantities
depending on the value of the wave function at
small distances. However, is
not clear the importance of this discrepancy
in the evaluation of the
phase shifts. From the technical point of view
a complete RGM calculation would be very
complicated to perform
with the full solution of the Schr\"{o}dinger
equation, but one can easily compare
the diagonal kernels of the NN potential
(which correspond to the diagonal part of those
used in Refs. \cite{FER,VAL} to
calculate the phase shifts and
the deuteron properties)
with both types of wave functions
as shown in Figure 3. We see how
the NN potential calculated with the full
solution of the Schr\"{o}dinger
equation looks pretty similar to the
potential calculated with the harmonic
oscillator wave function with
$b$ = 0.518 fm, this is precisely the
value used in the RGM calculation of
the phase shifts and
deuteron properties conferring to
it a self-consistent character.
Other values of $b$ give quite different
potentials even in the medium-long
range. In this way, the validity of
the gaussian ansatz wave function
in the RGM calculations is confirmed,
and the old controversy
about the possible values of the
parameter $b$ can be solved.

As a summary, we have analyzed the
baryonic low energy spectrum with a
potential model that includes, besides
the usual one-gluon exchange, pion and
sigma exchanges between quarks
as dictated by the breaking of chiral symmetry.
It is worth to notice that all the parameters
of the interaction are fixed to reproduce the
two nucleon phenomenology and none of them
is fixed to the spectrum.

The predicted baryon spectrum up
to 0.7 GeV of excitation energy is reasonable, and a
linear form of the confinement is preferred.
The pion interaction favors the correct
relative position of
the positive and negative parity states.
Baryon wave functions are also obtained
and compared to the ansatz
considered in previous studies of
the baryon spectrum,
nucleon-nucleon phase shifts and
deuteron properties.
The difference in the short range behavior
suggests the possible relevance of a reanalysis of the
one-body properties calculated with gaussian ansatzs.
Concerning the long-range part, the ansatz gaussian wave
function is valid for the study of the two
body properties and
we can figure out the value
of the harmonic oscillator
parameter, $b$, to be consistently
used in two-nucleon conventional
calculations.

Although much work is needed along
this direction (strange sector, decay
process, etc.) we think that the proposed
model could serve as a good
starting point to achieve a unified
description of the spectrum and the
baryon-baryon interaction.

\acknowledgements

This work has been partially supported by Direcci\'on
General de Investigaci\'on Cient\'{\i}fica y T\'ecnica
(DGICYT) under the Contract No. PB91-0119, by CICYT
grant AEN 93-0234 and by EU project ERBCHBICT941800.

\begin{figure}
\caption{ Relative energy nucleon and delta spectrum up to 0.7 GeV
excitation energy. The solid line corresponds to the predictions
of our model with a
linear confining potential. The shaded region, whose size stands for
the experimental uncertainty, represents the experimental
data \protect{\cite{PAR}}.}
\label{fig1}
\end{figure}

\begin{figure}
\caption{ Main radial component of the nucleon wave function in terms
of the hyperradius $\rho$. The solid line corresponds to the solution
of the Schr\"{o}dinger equation.
The long, medium and short-dashed lines to
gaussian ansatzs with values
of the parameter  $b = 0.4 \, , 0.518 \, , 0.6$ respectively.}
\label{fig2}
\end{figure}

\begin{figure}
\caption{ Diagonal kernels of the NN potential for
the channel (S,T)=(1,0) and for
different nucleon wave functions.
The solid line corresponds to the solution
of the Schr\"{o}dinger equation.
The long, medium, and short-dashed lines to
gaussian ansatzs with values
of the parameter  $b = 0.4 \, , 0.518 \, , 0.6$ respectively.}
\label{fig3}
\end{figure}

\begin{table}
\caption{ Quark model parameters.}
\label{quark}

\begin{tabular}{cccc}
 & $m_q (MeV)$                   &  313       & \\
 & $\alpha_s$                    &  0.485     & \\
 & $a_c (MeV \cdot fm^{-1})$     &  91.488    & \\
 & $\alpha_{ch}$                 &  0.0269    & \\
 & $r_0 (fm)$                    &  0.0367    & \\
 & $m_\sigma (fm^{-1})$          &  3.42      & \\
 & $m_\pi (fm^{-1})$             &  0.7       & \\
 & $\Lambda (fm^{-1})$           &  4.2       & \\
\end{tabular}
\end{table}

\begin{table}
\caption{ Relative energy (MeV) of the Roper and the first negative
parity nucleon excitations.}
\label{models}

\begin{tabular}{ccccc}
 &                                   & $N (1440)$ & $N^*(1535)$ & \\
\tableline
 & $V=V_{con}+V_{OGE}+V_{OPE}+V_{OSE}$ & 530 & 512  & \\
 & $V=V_{con}+V_{OGE}+V_{OSE}$         & 521 & 365  & \\
\end{tabular}
\end{table}

\end{document}